\documentclass[amssymb,prb,twocolumn,showpacs]{revtex4}
\usepackage{epsfig}
\usepackage{dcolumn}
\usepackage{amsmath}
\begin{document}
\title{\bf\ Microwave-induced resistance oscillations and
zero-resistance states in 2D electron systems with
two occupied subbands}
\author{J. I\~narrea$^{1}$
 and G. Platero$^2$}
\affiliation{$^1$Escuela Polit\'ecnica
Superior, Universidad Carlos III, Leganes, Madrid, 28911, Spain\\$^2$Instituto de Ciencia de Materiales,
CSIC, Cantoblanco, Madrid, 28049, Spain
}
\date{\today}
\begin{abstract}
We report on theoretical studies of recently discovered
microwave-induced resistance oscillations and zero resistance states in Hall bars with two occupied subbands.
In the same results, resistance presents a peculiar shape which appears
to have a built-in interference effect not observed before. We apply the microwave-driven electron orbit
model, which implies a radiation-driven oscillation
of the two-dimensional electron system. Thus, we calculate different intra and inter-subband electron scattering rates and times that are revealing as
different microwave-driven oscillations frequencies for the two electronic subbands. Through
scattering, these subband-dependent oscillation motions
interfere giving rise to a striking resistance profile. We also study the dependence of
irradiated magnetoresistance with power and temperature.
Calculated results are in good agreement with experiments.
\end{abstract}
\maketitle
\section{ Introduction}
Transport excited by radiation in a two-dimensional electron system (2DES) is
currently a central topic from experimental and theoretical
standpoints\cite{ina1}. The interest is focussed not only on the basic
explanation of a physical effect but also on its potential
applications.
In the last decade it was discovered that when a Hall bar (a 2DES
with a uniform and perpendicular magnetic field ($B$)) is irradiated
with microwaves, some unexpected effects are revealed, deserving
special attention from the condensed matter community:
microwave-induced (MW) resistance oscillations (MIRO) and zero
resistance states (ZRS) \cite{mani1,zudov1,studenikin1}. These
remarkable effects show up at low $B$ and high mobility samples,
specially ZRS where ultraclean samples are needed. Different
theories have been proposed to explain these striking effects
\cite{ina2,girvin,dietel,lei,ryzhii,rivera,shi} but the physical
origin is still being questioned. To shed some light on the physics
behind them, a great effort has been made, specially from the
experimental side, growing better samples, adding new features and
different probes to the basic experimental setup, etc.\cite{mani2,
mani3,willett,mani4,smet,yuan,stone,doro,hatke,mani5,mani6,wiedmann1,wiedmann2,kons1,kons2,vk}.
Of course the experimental results always mean a real challenge
for the existent theoretical models. Thus, a comparison of
experiment with theory could help to identify the importance of the
invoked-mechanisms in these theories.

One of the most interesting setups, carried out recently,
consists in using samples with two or three
occupied subbands\cite{wiedmann1}. These samples
are either based  in  a double quantum well structure or just one single but wide quantum well.
The main
difference in the longitudinal magnetoresistance ($R_{xx}$)
 of a two-subband sample is the presence of
magneto-intersubband oscillations (MISO)\cite{mamani}. These oscillations
occur due to periodic modulation of the probability of
transitions through elastic scattering between Landau levels (LL) of different subband.
The MISO peaks corresponds to the subband alignment condition
$\Delta=n\hbar w_{c}$, where $\Delta$ is the subband separation and $w_{c}$
the cyclotron frequency. Because of elastic scattering
of electrons between LL of different subband,
the probability rate is maximal under this condition.
Under
MW irradiation the first experimental results\cite{wiedmann2}
 of $R_{xx}$ showed the interference of MISO and MIRO
without reaching the ZRS regime. Later on, further experiments realized
at higher MW intensities and
mobility samples, showed  the MW-response to evolve into
  zero resistance states for the first time in
a two occupied subband sample\cite{wiedmann1}. In the same
experiment\cite{wiedmann1} it was also observed a peculiar $R_{xx}$
profile with different features, regarding the one-subband case
\cite{mani1,zudov1,studenikin1}, affecting only valleys and peaks
of MIRO's in a surprising regular way.
Thus, in  valleys
we observe two nearly symmetric shoulders, one at each
side of  minimum
which could correspond to a more intense
transport through the sample.
On the other hand, in the peaks we
observe narrower profiles, regarding again the one
subband case,
meaning a smaller transport.
\begin{figure}
\centering \epsfxsize=3.in \epsfysize=3.4in
\epsffile{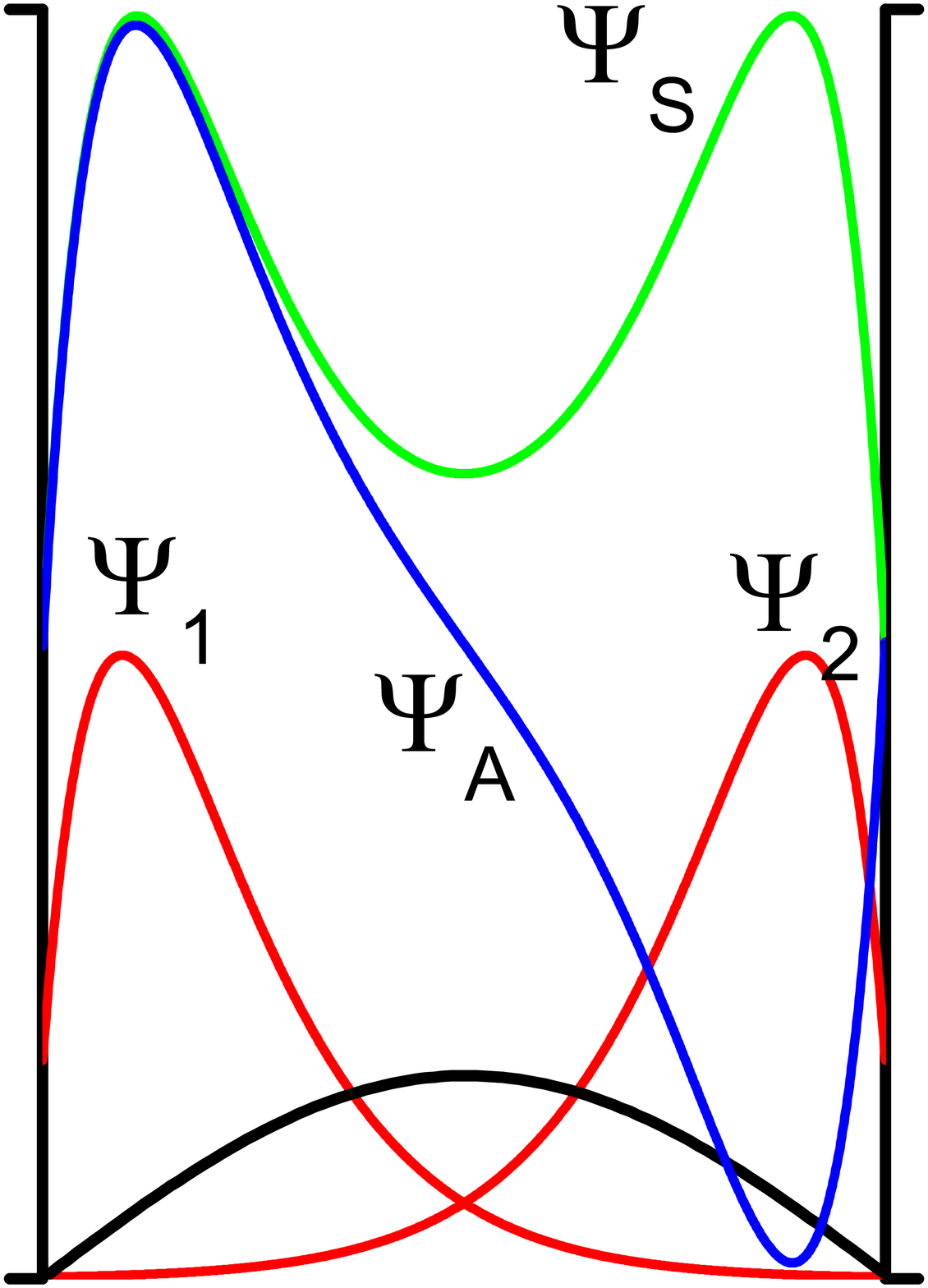}
\caption{Schematic diagram for the wide quantum well and the
corresponding electronic wave functions. $\Psi_{1}$ and $\Psi_{2}$
are the individual Fang-Howard wave functions for the left and right
triangular-like built-in potential wells. $\Psi_{S}$ and $\Psi_{A}$
are the symmetric and antisymmetric wave functions of the wide quantum well.
The quantum well has a  width of $45 nm$ as in the experiments.
}
\end{figure}

In this article, we
theoretically study magnetoresistance of a Hall bar being
illuminated with MW radiation when two electronic
subbands participate in the transport.  We apply
the theory developed by the authors, {\it the MW-driven electron
orbits model}\cite{ina2,ina3,kerner,park}, which we extend to a two-subband
scenario. According to this
theory, when a Hall bar is illuminated, the electron
orbit centers of the Landau states perform a classical trajectory consisting in a harmonic
motion along the direction of the current. Thus, the  2DES moves
periodically at the MW frequency altering dramatically the scattering
conditions and giving rise eventually to MIRO and ZRS.
In some cases the transport is reinforced producing
MW-induced $R_{xx}$ peaks; in others, transport is weakened giving
rise to valleys.
In a double subband scenario the situation gets more complicated but
with a richer physics. On the one hand, due to the presence of MW, we
have two 2DES (twoo subbands) moving harmonically at the
MW-frequency. On the other hand, we have two possible scattering
processes with charged impurities: intra and inter-subband. We then
calculate the two corresponding elastic impurity scattering rates,
obtaining that the intra is, approximately, three times larger
than the inter.
This means first, that the current is mainly supported by intra-subband scattering
processes.
Secondly and more important, the competition between intra and inter-subband
scattering events under the presence of radiation alters significantly the transport properties
of the sample. This is reflected in the $R_{xx}$ profile
through a strong and peculiar interference effect. As in experiments, our calculated results recover
the presence of new features regularly spaced through the
whole MIRO's profile, mainly two shoulders at minima  and narrower peaks.
We identify
such features with
situations where the interference is constructive and the current is
reinforced (shoulders around minima) meanwhile in other cases the
interference is destructive giving rise to a
less intense current (thinner peaks).
 Within the same theory, we
have obtained also ZRS in the same position of experiments and with
the same MW-frequency dependence. Finally, we have studied the influence of
MW-frequency ($w$), MW-power ($P$) and temperature ($T$) on MIRO's of the two subband sample and the
obtained results are also
in reasonable agreement with experiment\cite{wiedmann1}.


\section{ Theoretical Model}
The {\it MW driven electron orbits model}, was developed to explain
the $R_{xx}$ response of an irradiated 2DEG at low $B$. We first obtain
an exact expression of the electronic wave vector for a 2DES in a
perpendicular $B$, a DC electric field and MW radiation which is
considered semi-classically.
 Then, the total hamiltonian $H$ can be written as:
\begin{eqnarray}
H&=&\frac{P_{x}^{2}}{2m^{*}}+\frac{1}{2}m^{*}w_{c}^{2}(x-X)^{2}-eE_{dc}X +\nonumber \\
 & &+\frac{1}{2}m^{*}\frac{E_{dc}^{2}}{B^{2}}\nonumber-eE_{0}\cos wt (x-X) -\nonumber \\
 & &-eE_{0}\cos wt X \nonumber\\
 &=&H_{1}-eE_{0}\cos wt X
\end{eqnarray}
 $X$ is the center of the orbit for the electron spiral motion:
\begin{equation}
X=\frac{\hbar k_{y}}{eB}- \frac{eE_{dc}}{m^{*}w_{c}^{2}}
\end{equation}
$E_{0}$ the intensity for the MW field and $E_{dc}$ is the DC
electric field in the $x$ direction. $H_{1}$ is the hamiltonian
corresponding to a forced harmonic oscillator whose orbit is
centered at $X$. $H_{1}$ can be solved exactly \cite{kerner,park},
and using this result allows an exact solution for the electronic
wave function of $H$ to be obtained\cite{ina2,ina3,kerner,park,ina4}:
\begin{eqnarray}
\Psi_{N}(x,t)\propto\phi_{n}(x-X-x_{cl}(t),t)
\end{eqnarray}
where $\phi_{n}$ is the solution for the
Schr\"{o}dinger equation of the unforced quantum harmonic
oscillator, $x_{cl}(t)$ is the classical solution of a forced  and damped harmonic
oscillator
\begin{equation}
x_{cl}=\frac{e E_{o}}{m^{*}\sqrt{(w_{c}^{2}-w^{2})^{2}+\gamma^{4}}}\cos wt=A\cos wt
\end{equation}
where $\gamma$ is a phenomenologically-introduced damping factor
for the electronic interaction with acoustic phonons.

\begin{figure}
\centering\epsfxsize=3.5in \epsfysize=3.5in
\epsffile{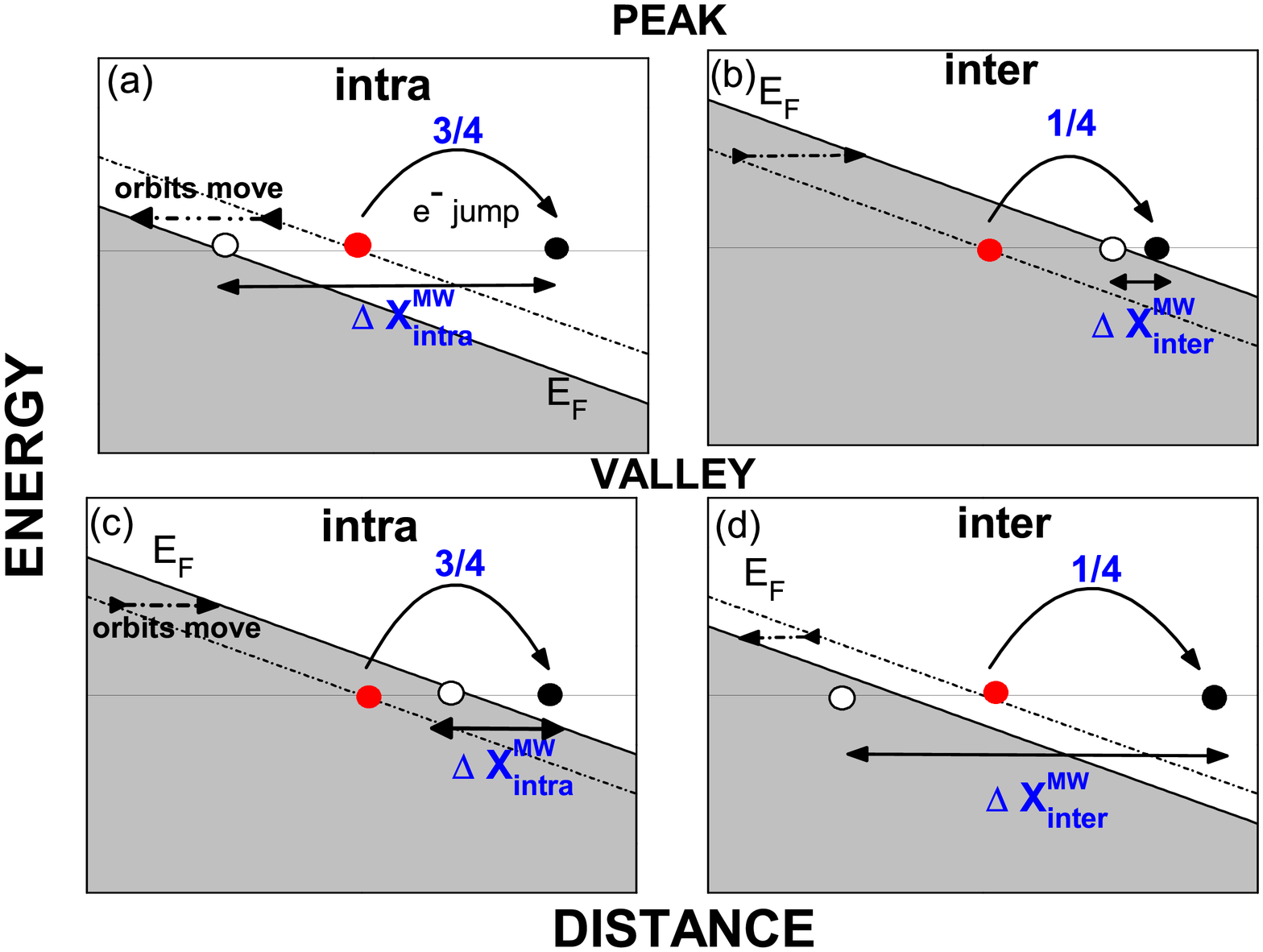}
\caption{Schematic diagrams of electronic transport corresponding to
peak and valley scenarios and intra and inter-subband scattering types.
When the
MW field is on, the orbits are not fixed but oscillate at $w$. In 2a,
in the intra-band processes corresponding to peaks, the electronic orbits are going
backward giving a larger advanced distance and current. Yet, only $3/4$ of
these processes develop through this channel.  In
2.b we represent the inter-subband  scattering processes for a peak, where due
to the lower scattering time the orbits are now going forwards, yielding
a smaller advanced distance and current.
 According to our calculations they are $1/3$ of the total.
Summing up all process, the total advance distance is
given by $\frac{3}{4}\Delta X^{MW}_{intra}+\frac{1}{4}\Delta X^{MW}_{inter}$.
Then, we obtain less current with respect to the one-subband case
due to the lower contributions of the inter-subband events.
This result  correspond to the obtained narrower
profiles at maxima.
The valleys situation reflected in 2c and 2d, can be explained in similar terms as the peaks but
now the discussion and results are going to be the opposite.
Then, considering jointly all
scattering processes, intra and inter-subband, we obtain, at both sides 
of minima,  more current than the
one-subband case due to larger contribution of the inter
scattering events. This is the physical origin of the two shoulder that
can be observed a both sides at minima.}
\end{figure}
Then, the obtained wave
function is the same as the standard harmonic oscillator where the
center is displaced by $x_{cl}(t)$.
Thus, the electron orbit centers are not
fixed, but they oscillate harmonically at $w$.
 This $radiation-driven$ behavior will affect dramatically the
charged impurity scattering and eventually the conductivity. Thus,
 we introduce the scattering suffered by the electrons due to
charged impurities.
 If the scattering is weak, we
can apply time dependent first order perturbation theory. First, we
calculate the impurity scattering rate\cite{ina2,ina3,ridley}
between two $oscillating$ Landau states $\Psi_{N}$,
final state $\Psi_{m}(x,t)$:
\begin{equation}
W_{n,m}=\lim_{\alpha\rightarrow 0} \frac{d}{d t} \left|
\frac{1}{i \hbar} \int_{-\infty}^{t^{'}}<\Psi_{m}(x,t) |V_{s}|\Psi_{n}(x,t)>e^{\alpha t}d t\right|^{2}
\end{equation}
where $V_{s}$ is the scattering potential for charged impurities\cite{ando}:
\begin{equation}
V_{s}= \sum_{q}\frac{e^{2}}{2 S \epsilon (q+q_{0})} \cdot e^{i
\overrightarrow{q}\cdot\overrightarrow{r}}
\end{equation}
$S$ being the surface of the sample, $\epsilon$ the GaAs dielectric
constant, and $q_{0}$ is the Thomas-Fermi screening
constant\cite{ando}.

After some lengthy algebra we arrive at the expressions for
the intra-subband
$W_{n,m}^{intra}$  and  the
inter-subband $W_{n,m}^{inter}$ scattering rates:
\begin{equation}
W_{n,m}^{intra}= |F_{intra}|^{2}\frac{e^{5}n_{i}S B m^{*} }
{\hbar^{4}\epsilon^{2}q_{0}^{2}}\left[1+2\sum_{s=1}^{\infty}e^{\left(\frac{-s\pi\Gamma}{\hbar w_{c}}\right)}\right]
\end{equation}
\\
\begin{eqnarray}
W_{n,m}^{inter}&=&|F_{inter}|^{2} \frac{e^{5}n_{i}S B m^{*} }
{\hbar^{4}\epsilon^{2}q_{0}^{2}}\times \nonumber\\
&&\left[1+2\sum_{s=1}^{\infty}
e^{ \left(\frac{-s\pi\Gamma}{\hbar
w_{c}}\right)}\cos\left(\frac{s2\pi\Delta_{12}}{\hbar
w_{c}}\right)\right]
\end{eqnarray}

where $n_{i}$ is the density of impurities, $\Gamma$ the width of the Landau states,
$\Delta_{12}$ the subband separation and
  $F_{S}$ and $F_{A}$ are the form
factors.
To obtain the form factor expressions we have considered, as in experiments\cite{wiedmann1},
a highly doped wide quantum well with $n_{i}\simeq 9.1\times 10^{11}cm^{}$. In this type of wells, as more
electrons are added, their electrostatic repulsion forces
them to pile up near the well sides and the resulting electron
charge distribution appears increasingly as bilayer (see Fig. 1). In other words,
and effective electrostatic barrier is built up in the middle of
the well separating the two
electron layers in GaAS. In the case of a double quantum dot the
barrier is made of ALGaAs or AlAs. As a result we obtain at each
side of the wide quantum well a potential profile similar to
the inversion layer of a 2DES that, in a good approximation,  can be considered
triangular close to the heterojunctions (see Fig. 1).
Next, we have to obtain first the corresponding
wave functions of these triangular-like built-in potential profiles.
Then, due to its great simplicity and as a first approach,
we have applied  the Fang-Howard variational treatment (see ref.\cite{ando,davies}) that
proposes
as electronic wave function (Fang-Howard wave function):
\begin{equation}
\Psi(z)=\left[\frac{b^{3}}{2}\right]^{1/2} z  e^{-\frac{1}{2}bz},
\end{equation}
 where $b$ is the corresponding Fang-Howard variational parameter.
According to this simple but efficient approach, $b$ results to be mainly
dependent on the two-dimensional charged impurity density\cite{davies,ando}.
Now starting from this variational wave function we can build $\Psi_{S(A)}$
which are the corresponding symmetric (antisymmetric) wave
function of the wide quantum well(see Fig. 1). Finally the form
factors are  obtained:
\begin{figure}
\centering \epsfxsize=3.5in \epsfysize=5.0in
\epsffile{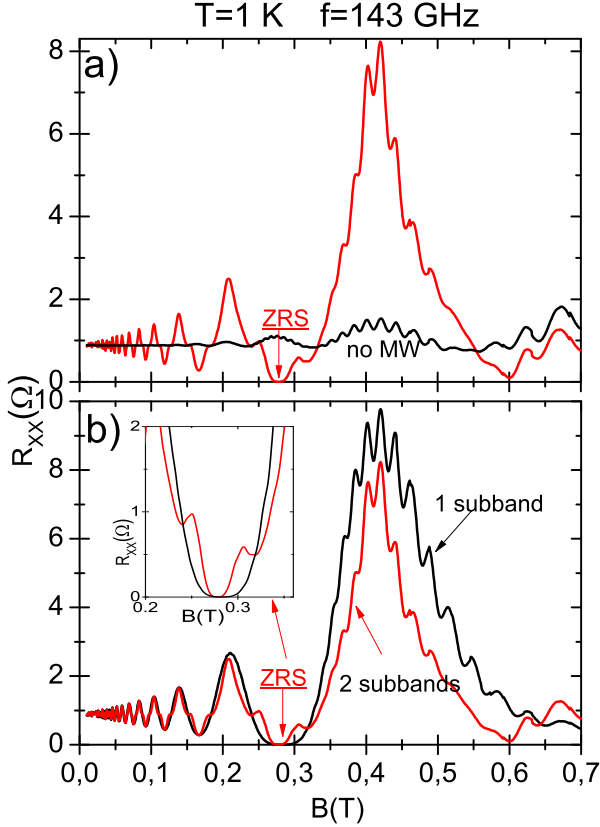}
\caption{a) Calculated $R_{xx}$ vs $B$ for dark and MW situations;
the ZRS is marked with an arrow.
b) Same as a) for $2$-subbands and $1$-subband.  We
observe clearly the new features showing up in the $2$-subband curve comparing to the
$1$-subband; shoulders at minima and narrower peaks.  In the inset the ZRS
region is blown up. Shoulders and narrower peaks are the outcomes of the interference
between the intra and inter-subband scattering processes.}
\end{figure}
\begin{widetext}
\begin{equation}
F_{intra}=\int_{0}^{\infty}e^{-q(z-z_{i})}\Psi_{S}^{*}\Psi_{S}  dz
=\frac{e^{-qd}}{2}\left[\left(\frac{b}{b+q}\right)^{3}+
\left(\frac{b}{b-q}\right)^{3}\right]
\end{equation}
\begin{equation}
F_{inter}=\int_{0}^{\infty}e^{-q(z-z_{i})}\Psi_{S}^{*}\Psi_{A} dz
=\frac{e^{-qd}}{2}\left[\left(\frac{b}{b+q}\right)^{3}-
\left(\frac{b}{b-q}\right)^{3}\right]
\end{equation}
\end{widetext}
where $q$ is the electron wave vector exchanged in the scattering. We have supposed a symmetrical
 delta doping, being $d$ the average separation between the impurities and the
2DES at each side of the wide quantum well, i.e., spacer distance. This distance
depends on the sample. In practical terms it varies between
$10 nm$ to even larger than $100 nm$. In our calculations
we have used a numerical value of $d=70-90 nm$. According to the
Fang-Howad variational approach\cite{davies,ando}, where the
 parameter $b$ is given by:
\begin{equation}
b=\left[\frac{33m^{*}n_{i}}{8\hbar^{2}\epsilon}\right]^{1/3}
\end{equation}
, and applying the experimental sample parameters\cite{wiedmann1},
we have calculated a numerical value for $b\approx 0.25 nm^{-1}$. Consequently the average thickness of the triangular wells $\langle z \rangle \simeq 11-12 nm$, where $\langle z \rangle$ is  related with $b$ by  $\langle z \rangle=\frac{3}{b}$.
Following again with the experimental
parameters at hand\cite{wiedmann1} in terms of impurity density, well
thickness, etc., we have made an averaged estimation of the
relative values of $F_{S}$ and $F_{A}$ resulting in
\begin{equation}
|F_{S}|^{2}=3.2 \times|F_{A}|^{2}
\end{equation}
where we have used an average value
for $\overline{q}=0.5 nm^{-1}$ that is of the order of the
Fermi wave vector in agreement with the experimental impurity density.
Substituting the obtained form factors in the scattering rates
we can eventually reach averaged values for those rates that
result to be related by:
\begin{equation}
\langle
W_{n,m}^{intra}\rangle \approx 3\times \langle
W_{n,m}^{inter}\rangle
\end{equation}
Here we have considered that
the cosine average value,
$\left\langle\cos\frac{s2\pi\Delta_{12}}{\hbar
w_{c}}\right\rangle\rightarrow0$
for $\Delta_{12} >\hbar w_{c} $ and  we have carried out the sum
$\sum_{s=1}^{\infty}e^{\left(\frac{-s\pi\Gamma}{\hbar
w_{c}}\right)}\rightarrow \frac{exp(\frac{-s\pi\Gamma}{\hbar
w_{c}})}{1-exp(\frac{-s\pi\Gamma}{\hbar
w_{c}})}$.

Once we know the intra and inter-subband scattering rates, we consider that  when
an electron undergoes a scattering process jumping from the initial state
to the final one,  it takes an
average time
\begin{equation}
\langle\tau_{intra(inter)}\rangle=\left\langle\frac{1}{W_{n,m}^{intra(inter)}}\right\rangle
\end{equation}
Following the model described in ref.\cite{ina2},  we next find the average effective distance advanced by the electron
in every scattering jump in the presence of radiation $\Delta X^{MW}$ , generalizing the previous results to
a two subbands scenario:
\begin{equation}
\Delta X^{MW}_{intra(inter)}=\Delta X^{0}+ A\cos
(w\langle\tau_{intra(inter)}\rangle)
\end{equation}
 where $\Delta X^{0}$ is the
effective distance advanced when there is no MW field present.
Applying the important previous result of
\begin{equation}
\langle
W_{n,m}^{intra}\rangle \approx 3\times \langle
W_{n,m}^{inter}\rangle\Rightarrow \langle \tau_{intra}\rangle\approx
\frac{1}{3} \langle \tau_{inter}\rangle
\end{equation}
we can
write the final expression for the total average distance advance due to
both kinds of scattering, intra and inter, $\Delta X^{MW}_{total}$:
\begin{eqnarray}
\Delta X^{MW}_{total}&=&\Delta X^{MW}_{intra}+\Delta X^{MW}_{inter}\\
&=& A\cos\left[w \langle\tau_{intra}\rangle\right]+
A\cos \left[w\langle\tau_{inter}\rangle\right]\\
&=&A\cos\left[\frac{w}{3}\langle\tau_{inter}\rangle\right]+
A\cos \left[w\langle\tau_{inter}\rangle\right]
\end{eqnarray}
This significantly alters the scattering
conditions regarding the one-subband case mainly affecting MIRO's peaks and valleys.
\begin{figure}
\centering \epsfxsize=3.5in \epsfysize=5.0in
\epsffile{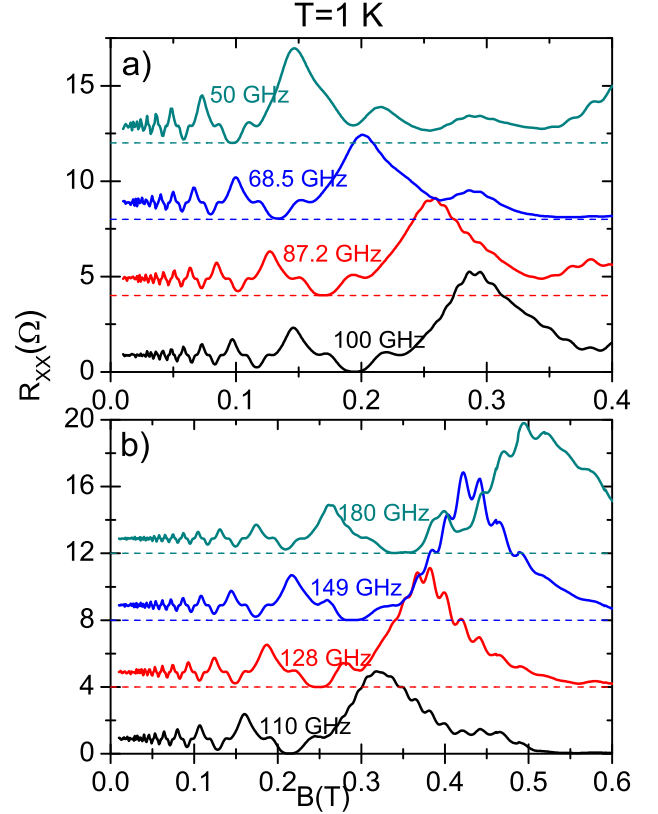}
\caption{Calculated $R_{xx}$ vs $B$ for different frequencies. In a) for lower
frequencies, from $50$ to $100$ GHZ, and in b) for higher, from $110$ to $180$ GHZ.
We observe the presence of ZRS in all curves with shifting
positions depending on the frequency and the interference features in peaks and valleys.
ZRS positions and shifts are in agreement with
experiments\cite{wiedmann1,wiedmann3} and are similar to the one-subband case.}
\end{figure}

Finally the contributions intra and inter-subband to the longitudinal
conductivity can be calculated: $\sigma_{xx}\propto
\int dE [(\frac{\Delta X^{MW}}{\tau})_{intra}+(\frac{\Delta X^{MW}}{\tau})_{inter}](f_{i}-f_{f})$,  being $f_{i}$ and
$f_{f}$ the corresponding distribution functions for the initial and
final Landau states respectively and $E$ energy. The obtained final expression
for the conductivity is given by:
\begin{widetext}
\begin{eqnarray}
&&\sigma_{xx}=\frac{6e^{7}m^{*2}B n_{i}S}{\pi\epsilon^{2}\hbar
 ^{6}q_{0}}\left[\Delta X^{0}+ A\cos
\frac{1}{3}w\langle\tau_{inter}\rangle\right]^{2}
\left[1+2e^{\frac{-\pi\Gamma}{\hbar w_{c}}}+e^{\frac{-\pi\Gamma}
{\hbar w_{c}}}\frac{X_{S}}{\sinh X_{S}}\left(\cos\frac{2\pi(E_{F}-E_{1})}{\hbar w_{c}}+\cos \frac{2\pi(E_{F}-E_{2})}{\hbar w_{c}}\right)\right]\nonumber\\
&&+\frac{2e^{7}m^{*2}B n_{i}S}{\pi\epsilon^{2}\hbar
^{6}q_{0}}\left[\Delta X^{0}+ A\cos
w\langle\tau_{inter}\rangle \right]^{2}
\left[1+2e^{\frac{-\pi\Gamma}{\hbar w_{c}}}\cos\frac{2\pi\Delta_{12}}{\hbar w_{c}}+e^{\frac{-\pi\Gamma}
{\hbar w_{c}}}\frac{X_{S}}{\sinh X_{S}}\left(\cos\frac{2\pi(E_{F}-E_{1})}{\hbar w_{c}}+\cos\frac{2\pi(E_{F}-E_{2})}{\hbar w_{c}}\right)\right]\nonumber\\
\end{eqnarray}
\end{widetext}
\begin{figure}
\centering \epsfxsize=3.5in \epsfysize=5.0in
\epsffile{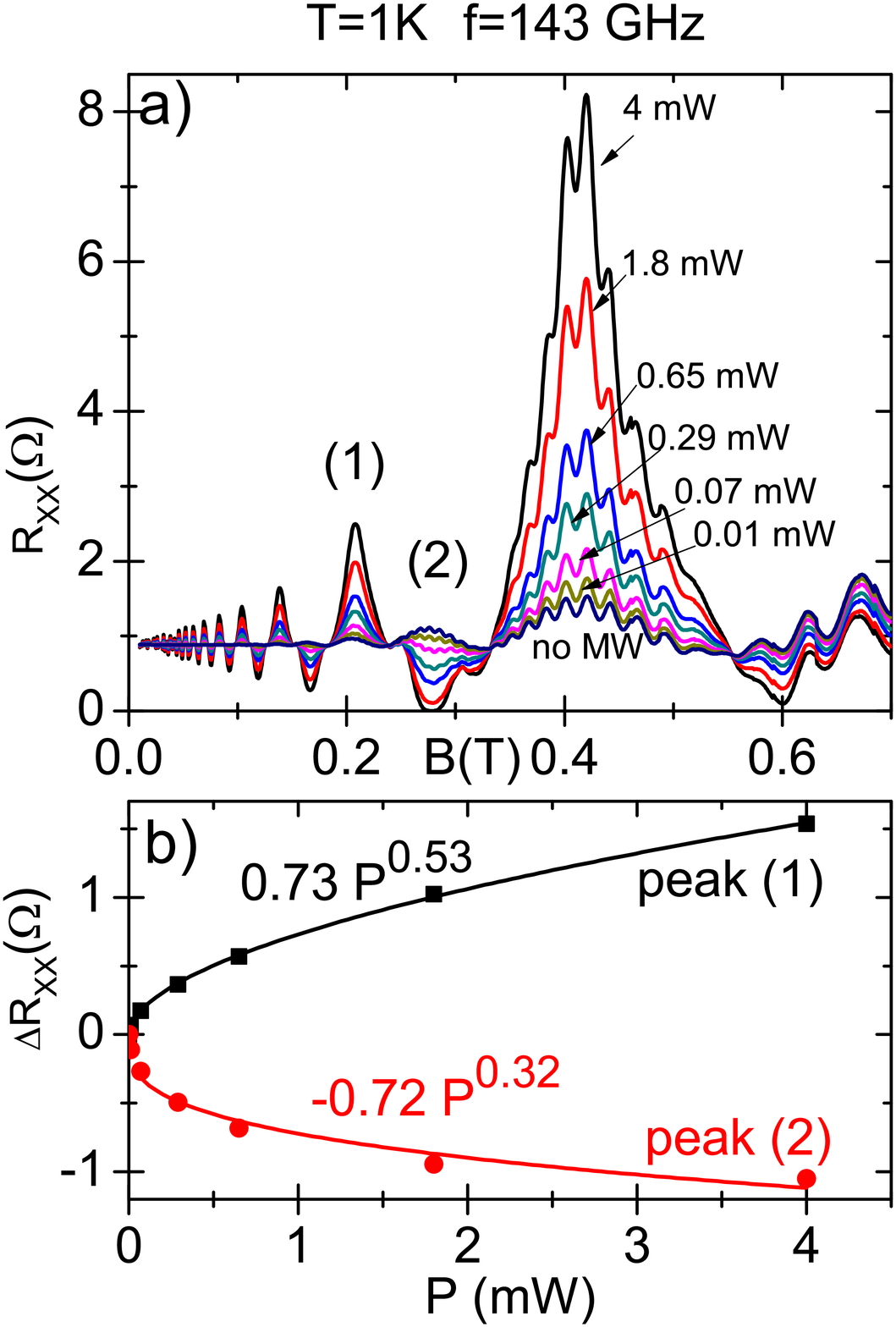}
\caption{ a) Calculated results of power dependence of $R_{xx}$ vs $B$. MW power decreases from $4$ mW to
darkness and  MIRO's decrease too. This is a similar behavior as
the $1$-subband result. b) Calculated $\Delta R_{xx}=R_{xx}^{MW}-R_{xx}^{0}$ vs power, for
data coming from peaks (1) and (2) of the upper panel. The obtained fits for both
peaks mean a sublinear $P$-dependence in agreement with previous
experimental\cite{mani6} and theoretical\cite{ina5} results.}
\end{figure}
\begin{figure}
\centering \epsfxsize=3.5in \epsfysize=5.0in
\epsffile{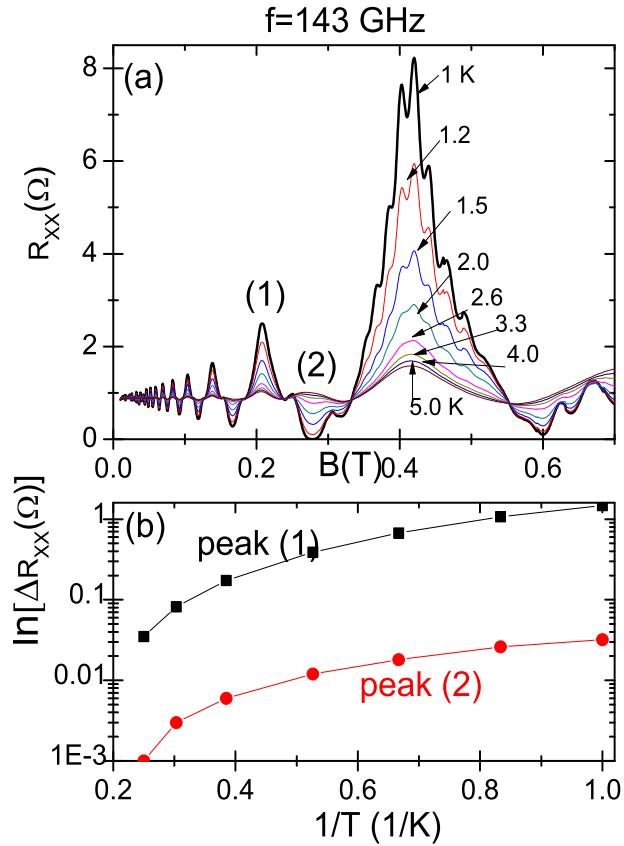}
\caption{ a) Calculated results of $T$ dependence of $R_{xx}$ vs $B$
for $f=143$ GHz. As in $1$-subband samples, we observe a clear decrease of MIRO for increasing $T$,
eventually reaching a $R_{xx}$ response similar to darkness. b)  $ln\Delta R_{xx}$ vs $1/T$ for data coming from peaks (1) and (2) of 6a. The fitted curves show the  relation $R_{xx}\propto T^{-2}$ in agreement with experiment.
The $T$-dependence is explained
with the damping parameter $\gamma$ which represents the interaction of electrons with
acoustic phonons. $\gamma$ is linear with $T$\cite{ina2,ina3}, thus an increasing $T$ means
an increasing $\gamma$ and smaller MIRO's. When the damping is strong enough (higher $T$), MIRO's collapse}
\end{figure}

where $X_{S}=\frac{2\pi^{2}k_{B}T}{\hbar w_{c}}$ and $E_{1}$ and $E_{2}$ are
the energies of the first and the second subband respectively.
Finally, to obtain
$R_{xx}$ we use the relation
$R_{xx}=\frac{\sigma_{xx}}{\sigma_{xx}^{2}+\sigma_{xy}^{2}}
\simeq\frac{\sigma_{xx}}{\sigma_{xy}^{2}}$, where
$\sigma_{xy}\simeq\frac{n_{i}e}{B}$ and
$\sigma_{xx}\ll\sigma_{xy}$.

The expression of the conductivity $\sigma_{xx}$ shows the physical equivalence to a situation with
only one scattering time and two different oscillations frequencies
for the MW-driven subbands: $w/3$ for the intra-band scattering and
$w$ for the inter. They demonstrate also the origin for the regular
and strong interference profile
observed in experiments where the factor $1/3$ is essential
to obtain the interference effect.
A different factor would produce a totally distinct interference and also
distinct $R_{xx}$ response. This factor comes from the calculation
of the squared magnitude of the corresponding form factors $F_{intra}$ and $F_{intra}$
which eventually determine the different scattering rate between the intra-subband
and the inter-one processes.
These form factors  depend mainly on the variational parameter $b$ and on the
averaged wave vector $\overline{q}$ which subsequently are determined by
two-dimensional impurity density $n_{i}$. Therefore we can conclude
that the crucial parameter $1/3$ and eventually the obtained
interference profile will be mainly dependent on $n_{i}$. During the scattering jump the electron {\it perceives}
an approximately three times faster MW-driven oscillation of the 2DES when is
inter-subband with respect to the intra-subband.
This equation reflects also the important result that the intra-band conductivity (upper
term in the $\sigma_{xx}$ expression) is
three times larger than the inter-subband (lower term of $\sigma_{xx}$). Then, the total current is mainly
supported by intra-subband scattering processes regarding the inter processes in a relation of
approximately three to one.

Based in these results, we can explain physically how the interference
between both types of scattering process work, producing an excess of current at
minima and a lack of current at maxima. In Fig. 2 we present schematic
diagrams for the different situations.  In all of them the MW field is on and
the electronic orbits are not fixed, and instead
move back and forth through $x_{cl}$. In Fig. 2a, the intra-subband
scattering corresponds to a peak and the orbits moves backwards
during the jump, then on average, electrons advance further than in the
no MW case. Thus, we obtain more current giving rise to peaks
because the average advanced distance is directly proportional to
the conductivity. Yet, according
to our calculations, only three out of four of the total scattering
processes take place through the intra-subband channel. The
other one out of four are inter-subband processes (see Fig. 2b). In them, due to the slower scattering
time (or higher $w$) the scattering jump occurs when the electronic orbits
are moving forward. Therefore, the average advanced distance is
smaller than in the intra processes. Eventually at the peaks, the total advanced distance,
$\frac{3}{4}\Delta X^{MW}_{intra}+\frac{1}{4}\Delta X^{MW}_{inter}$,
is smaller than the one-subband case.
Accordingly, this is directly translated to  the obtained current, being
reflected in the narrower peaks profile (see Fig. 3b).

The valleys situation can be explained in similar terms as the peaks.
Now, in the intra processes (see Fig. 2c) the scattering jump takes place when the electronic
orbits are moving forward giving a smaller electronic advanced distance and
 current. This situation gives rise to
valleys and eventually if the MW power is big enough the ZRS can be
achieved. As  before, these intra-subband scattering processes correspond
to three out of four of the total (see Fig. 2c). The other one out
of four develops through the inter-subband channel (see Fig. 2d). In these processes
,again due to the smaller scattering time or higher $w$,  the scattering
jump occurs when the electronic orbits are going backwards giving rise to
a larger average advanced distance and  current (see Fig. 2d).
 Then, summing up all
scattering processes we obtain more current than the
one-subband case.
 This is the physical origin of the two shoulder that
can be observed a both sides at minima of $R_{xx}$ versus $b$.

\section{Results}
In Fig. 3a, we present calculated $R_{xx}$ vs $B$ for dark and MW situations
and frequency $f=w/2\pi=143$ GHz. We can observe MISO for the no-MW
curve, MIRO for the MW curve and the ZRS marked with and arrow. In Fig. 3b, we
present the same as in 3a, but for $2$-subbands and $1$-subband cases. The later
has been obtained making $\Delta_{12}\rightarrow0$. Contrasting both curves we observe
the new features appearing regularly spaced in peaks and valleys: two nearly
symmetric shoulders in valleys and narrower peaks regarding the $1$-subband curve.
According to our model, these new features are result of the interference between
the competing intra and inter-subband scattering processes. This interference
effect is mainly based in the different scattering rates between intra and
inter-subband scattering events. On the one hand this is going to be obviously
reflected in the different scattering times: the inter process is
three times smaller than the intra-subband. On the other hand, in
the different capability to support the current. The
intra-subband processes are able to support three times more current than the inter.
Thus, in valleys, we observe
a constructive interference
effect giving rise to two shoulders meaning more current through the sample,
meanwhile the narrower peaks mean a destructive interference and
 less current. The presence of ZRS is explained similarly
as in ref.\cite{ina2}.

In Fig.4 we present calculated $R_{xx}$ vs $B$ for different MW frequencies. In 4a, for a lower
frequencies range (from $50$ to $100$ GHZ)\cite{wiedmann3} and in 4b, for a
higher frequencies range (from $110$ to $180$ GHZ)\cite{wiedmann1}.
We observe in all cases the presence of ZRS with shifting
position depending on $f$ and with reasonable agreement with experiment. All curves
present the peaks and valley features meaning the importance of the interference
effect which shows up  independently of $f$. This $f$-dependence of ZRS positions
was previously and similarly observed in one-subband samples\cite{mani1,zudov1}.

In Fig. 5a we present $P$-dependence of $R_{xx}$ vs $B$ for $f=143$ GHz.
We observe that MIRO's decrease as $P$, (and $E_{0}$), gets smaller from $4$ mW to darkness in similar behavior as
the $1$-subband results\cite{mani1}. In 5b, we present  $\Delta R_{xx}=R_{xx}^{MW}-R_{xx}^{0}$ vs $P$, for
data coming from peaks (1) and (2) of Fig. 5a, where $R_{xx}^{0}$ is the magnetoresistance
for darkness and $R_{xx}^{MW}$ is the magnetoresistance
when the MW field is on. We fit the data obtaining for both peaks a sublinear
$P$-dependence, $R_{xx}\propto P^{\alpha}$ where $\alpha<1$ and explained in terms of:
\begin{equation}
 E_{0}\propto \sqrt{P}\Rightarrow R_{xx}\propto \sqrt{P}
\end{equation}
  and in agreement with current\cite{wiedmann1}
and previous
experimental\cite{mani6} and theoretical\cite{ina5} results.

 In Fig 6a. we present the $T$ dependence of $R_{xx}$ vs $B$
for $f=143$ GHz. As in $1$-subband samples, we observe a clear decrease of MIRO for increasing $T$,
eventually reaching a $R_{xx}$ response similar to darkness. In Fig. 6b, we present $ln\Delta R_{xx}$ vs $1/T$ for data coming from peaks (1) and (2) of Fig. 6a. The $T$-dependence, according to the model, is explained
with the damping parameter $\gamma$ which represents the interaction of electrons with
acoustic phonons. $\gamma$ is linear with $T$\cite{ina2,ina3}, thus an increasing $T$ means
an increasing $\gamma$ and smaller MIRO's. When the damping is strong enough (higher $T$) MIRO's collapse.
The curves of Fig. 6b show the relation $R_{xx}\propto T^{-2}$ in agreement with experiment\cite{wiedmann1}.
\\

\section{Conclusions}
In summary, we have theoretically studied the  recently discovered
microwave-induced resistance oscillations and zero resistance states in Hall bars with two occupied subbands.
MW-driven magnetoresistance presents a peculiar shape which appears
to have a built-in interference effect not observed before. Applying the microwave-driven electron orbit
model, we calculate different intra and inter-subband electron scattering rates under MW, revealing
that the first is three times greater
than the latter. This is physically equivalent to different microwave-driven oscillations frequencies for the two electronic subbands. Through
scattering, these subband-dependent oscillation motions
interfere giving rise to such a striking resistance profile. In the valleys
the interference is constructive giving rise to two symmetric
extra shoulders at each side of minima. In the peaks the
interference is destructive producing smaller peaks regarding
the one-subband case. The factor $\sim3$ is essential
to obtain this strong and regularly spaced interference effect.
We study also the dependence on
MW frequency, MW  intensity and temperature.
Calculated results are in good agreement with experiments.
\\
\\

\section{Acknowledgments}
This work is supported by the MCYT (Spain) under grant:
MAT2008-02626/NAN.


\begin{thebibliography}{19}

\section{References}
\bibitem{ina1}
J. I\~narrea, G. Platero and C. Tejedor, Semicond. Sci. Tech. {\bf
9}, 515, (1994);J. I\~narrea, G. Platero, Phys. Rew. B, {\bf 51}, 5244, (1995);Europhys. Lett.
{\bf 34}, 43, (1996);Europhys Lett.  {\bf 33}, 477, (1996); Europhys
Lett.  {\bf 40}, 417, (1997).

\bibitem{mani1} R. G. Mani, J. H. Smet, K. von Klitzing, V. Narayanamurti,
W. B. Johnson, and V. Umansky, Nature(London) \textbf{420}, 646
(2002); R. G. Mani, V. Narayanamurti, K. von Klitzing, J. H. Smet,
W. B. Johnson, and V. Umansky, Phys. Rev. B\textbf{69}, 161306
(2004); Phys. Rev. B\textbf{70}, 153310 (2004).

\bibitem{zudov1} M. A. Zudov, R. R. Du, L. N. Pfeiffer, and K. W. West,
Phys.
Rev. Lett. \textbf{90}, 046807 (2003).

\bibitem{studenikin1} S. A. Studenikin et al., Sol. St. Comm. \textbf{129}, 341
(2004).

\bibitem{ina2}
J. I\~narrea and G. Platero, Phys. Rev. Lett. {\bf 94} 016806,
(2005); J. I\~narrea and G. Platero, Phys. Rev. B {\bf 72} 193414
(2005);J. I\~narrea and G. Platero, Appl. Phys. Lett.,  {\bf 89},
052109, (2006);J. I\~narrea and G. Platero, Phys. Rev. B,  {\bf 76},
073311, (2007);  J. I\~narrea, Appl. Phys. Lett. {\bf 90}, 172118,
(2007)

\bibitem{girvin}
A.C. Durst, S. Sachdev, N. Read, S.M. Girvin, Phys. Rev. Lett.{\bf
91} 086803 (2003)

\bibitem{dietel}
C.Joas, J.Dietel and F. von Oppen, Phys. Rev. B {\bf 72}, 165323,
(2005)

\bibitem{lei}
X.L. Lei, S.Y. Liu, Phys. Rev. Lett.{\bf 91}, 226805 (2003)

\bibitem{ryzhii}
Ryzhii et al, Sov. Phys. Semicond. 20, 1299, (1986)

\bibitem{rivera}
P.H. Rivera and P.A. Schulz, Phys. Rev. B {\bf 70} 075314 (2004)

\bibitem{shi}
Junren Shi and X.C. Xie, Phys. Rev. Lett. {\bf 91}, 086801 (2003)



\bibitem{mani2}R. G. Mani et al., Phys. Rev. Lett. \textbf{92}, 146801
(2004).

\bibitem{mani3} R. G. Mani et al., Phys. Rev. B\textbf{69}, 193304
(2004).

\bibitem{willett} R. L. Willett, L. N. Pfeiffer, and K. W. West, Phys.
Rev. Lett. \textbf{93}, 026604 (2004).

\bibitem{mani4}R. G. Mani, Physica E (Amsterdam) \textbf{22}, 1 (2004);

\bibitem{smet} J. H. Smet et al., Phys. Rev. Lett. 95, 118604 (2005).

\bibitem{yuan} Z. Q. Yuan et al., Phys. Rev. B\textbf{74},
075313 (2006).

\bibitem{stone} K. Stone et al., Phys. Rev. B\textbf{76}, 153306 (2007).

\bibitem {doro} S. I. Dorozhkin et al., Phys. Rev. Lett. 102, 036602
(2009).

\bibitem{hatke} A. T. Hatke et al., Phys. Rev. Lett. \textbf{102}, 086808
(2009).

\bibitem{mani5} R. G. Mani et al., Phys. Rev. B\textbf{79}, 205320
(2009).

\bibitem{mani6} R. G. Mani et al., Phys. Rev. B\textbf{81}, 125320, (2010).

\bibitem{wiedmann1} S. Wiedmann, G.M. Gusev, O.E. Raichev, A.K. Bakarov, and
J.C. Portal, Phys. Rev. Lett., {\bf 105}, 026804, (2010)

\bibitem{wiedmann2} S. Wiedmann, G.M. Gusev, O.E. Raichev, A.K. Bakarov, and
J.C. Portal, Phys. Rev. B, {\bf 81}, 085311, (2010);
S. Wiedmann, N.C. Mamani, G.M. Gusev, O.E. Raichev, A.K. Bakarov, and
J.C. Portal, Phys. Rev. B, {\bf 80}, 245306, (2009);
S. Wiedmann, G.M. Gusev, O.E. Raichev, T.E. Lamas, A.K. Bakarov, and
J.C. Portal, Phys. Rev. B, {\bf 78}, 121301, (2008);

\bibitem{kons1}
D. Konstantinov and K. Kono, Phys. Rev. Lett. {\bf 103}, 266808
(2009)

\bibitem{kons2}
D. Konstantinov and K. Kono, Phys. Rev. Lett. {\bf 105}, 226801
(2010)

\bibitem{vk}
S. I. Dorozhkin, L. Pfeiffer, K. West K, K. von Klitzing, J.H. Smet JH,
NATURE PHYSICS, {\bf 7}, 336-341, (2011)

\bibitem{mamani}
O. E. Raichev, Phys. Rev. B 78, 125304 (2008);
N. C. Mamani, G. M. Gusev, O. E. Raichev, T. E. Lamas,
and A. K. Bakarov, Phys. Rev. B 80, 075308 (2009).

\bibitem{mani7} R. G. Mani, Int. J. Mod. Phys. B, {\bf 18}, 3473,
(2004); Physica E, \textbf{25}, 189 (2004)

\bibitem{zudov2} M. A. Zudov, R. R. Du, L. N. Pfeiffer, and K. W. West, Phys. Rev. Lett.
\textbf{96}, 236804 (2006)

\bibitem{kunold} Phys. stat. sol. (a), \textbf{204}, 467, (2007)

\bibitem{ina3} J. Inarrea and G. Platero, Appl. Physl Lett. {\bf 89},
172114, (2006)

\bibitem{kerner}
E.H. Kerner, Can. J. Phys. {\bf 36}, 371 (1958) .

\bibitem{park}
K. Park, Phys. Rev. B {\bf 69} 201301(R) (2004).

\bibitem{ina4}
J. I\~narrea and G. Platero, Appl. Phys Lett. {\bf 93}, 062104,
(2008); J. I\~narrea and G. Platero, Phys. Rev. B,. {\bf 78},
193310,(2008);J. I\~narrea, Appl. Phys Lett. {\bf 92},
192113,(2008); Jesus Inarrea and Gloria Platero, Appl. Phys. Lett.
{\bf 95}, 162106, (2008);J. I\~narrea, G. Platero and C. Tejedor, Semicond. Sci. Tech. {\bf
9}, 515, (1994);J. I\~narrea, G. Platero, Phys. Rew. B, {\bf 51}, 5244, (1995);Europhys. Lett.
{\bf 34}, 43, (1996);Europhys Lett.  {\bf 33}, 477, (1996); Europhys
Lett.  {\bf 40}, 417, (1997).

\bibitem{ridley}
B.K. Ridley. Quantum Processes in Semiconductors, 4th ed. Oxford
University Press, (1993).

\bibitem{ando}
T. Ando, A. Fowler and F. Stern, Rev. Mod. Phys.,{\bf 54},(1982).

\bibitem{davies}
John H. Davies, The Physics of Low-dimensional Semiconductors, Cambridge
University Press, (1997).

\bibitem{mitin}
V. Mitin, V.A. Kochelap and M.A. Stroscio. Quantum Heterostructures, Cambridge University
Press, (1999).
\bibitem{wiedmann3}
S. Wiedmann, G. M. Gusev, O. E. Raichev, A. K. Bakarov, and J. C. Portal, Proceedings of
the 19th International Conference on the Application of High Magnetic Fields in Semiconductor
Physics(HMF-19), Fukuoka, Japan, 2010 (unpublished)



\bibitem{ina5}
Jesus Inarrea, R.G. Mani and W. Wegscheider, Phys. Rev. ,{\bf 82} 205321 (2010);
R. G. Mani, C. Gerl, S. Schmult, W. Wegscheider, V. Umansky,
Phys. Rev. B  {\bf 81}, 125320 (2010).




\end{thebibliography}
\end{document}